**Machine learning models show similar performance to Renewables.ninja for generation of long-term wind power time series even without location information**

Baumgartner, J.[1], Gruber, K.[1], Simoes, S.G.[2], Saint-Drenan, Y.M.[3], Schmidt, J.[1]

1 Institute for Sustainable Economic Development, University of Natural Resources and Life Sciences, Vienna, Austria

2 CENSE – Center for Environmental and Sustainability Research, NOVA School of Science and Technology - NOVA University Lisbon, Portugal

3 MINES ParisTech, PSL Research University, O.I.E. Centre Observation, Impacts, Energy, 06904 Sophia Antipolis, France

## Abstract

Driven by climatic processes, wind power generation is inherently variable. Long-term simulated wind power time series are therefore an essential component for understanding the temporal availability of wind power and its integration into future renewable energy systems. In the recent past, mainly power curve based models such as Renewables.ninja (RN) have been used for deriving synthetic time series for wind power generation despite their need for accurate location information as well as for bias correction, and their insufficient replication of extreme events and short-term power ramps. We assess how time series generated by machine learning models (MLM) compare to RN in terms of their ability to replicate the characteristics of observed nationally aggregated wind power generation for Germany. Hence, we apply neural networks to one MERRA2 reanalysis wind speed input dataset with no location information and one with basic location information. The resulting time series and the RN time series are compared with actual generation. Both MLM time series feature equal or even better time series quality than RN depending on the characteristics considered. We conclude that MLM models can, even when reducing information on turbine locations and turbine types, produce time series of at least equal quality to RN.

# Introduction

Globally installed wind power capacity increased more than sixfold within eleven years from 93.9 GW in 2007 to 591.5.8 GW in 2018. As of 2018, nearly one tenth of global installed wind power capacity is located in Germany, where installed wind power capacity increased from 22.2 GW in 2007 to 53.2 GW in 2018. This increase resulted in a power generation of 111.5 TWh in 2018 corresponding to 21% of electricity demand in Germany (Arbeitsgemeinschaft Energiebilanzen, 2019; Global Wind Energy Council, 2019, 2008; WindEurope, 2019).

Due to this significant expansion, the spatial and temporal availability of climate dependent wind resources increasingly affects the whole power system. Consequently, assessments of the electricity system's vulnerability to climate extreme events and climate change by means of power system models provide vital insights on how future electricity systems should be structured to mitigate supply scarcity and power outages (Bonjean Stanton et al., 2016; Klein et al., 2013; Tobin et al., 2018). Therefore, accurate multi-year generation time series (i.e. multiple years of temporally highly resolved values) are used as input data to power (e.g. Réseau de transport d'électricité, 2018; Strbac et al., n.d.) as well as energy system models (e.g. E3M, 2018; Loulou et al., 2016; Simoes et al., 2017) for quantifying the system's resilience to climate events. This is increasingly important when higher market penetration of renewables is taken into account (Staffell and Pfenninger, 2016).

In the recent past, mainly power curve based models based on reanalysis climate data sets (Andresen et al., 2015; Cannon et al., 2015; Gruber et al., 2019; Staffell and Green, 2014; Staffell and Pfenninger, 2016; Zucker et al., 2016) have been used for deriving multi-year time series. These models have been able to reproduce wind power generation time series sufficiently well in terms of error metrics, distributional and seasonal characteristics. However, these models also feature possible drawbacks of high data needs for model setup (i.e. wind turbine locations, turbine specifications and commissioning dates) and the need for separate work steps for bias correction and for the replication of wake effects (e.g. power curve smoothing) (Olauson and Bergkvist, 2015; Staffell and Pfenninger, 2016; Zucker et al., 2016). In particular, well known shortcomings of reanalysis data, i.e. a significant mean bias in wind speeds, have to be overcome by the models via bias correction (Olauson and Bergkvist, 2015), which relies on the availability of historical wind power generation time series or independent sources of wind speed data, such as local wind speed measurements. Another downside of reanalysis-based time series generated by power curve based models has been their insufficient replication of extreme generation events as well as short-term power ramps (Cannon et al., 2015; Staffell and Pfenninger, 2016; Zucker et al., 2016). This can only partly be attributed to the methodology as the underlying reanalysis data sources do not sufficiently capture extreme situations (Zucker et al., 2016). The accurate replication of power generation extremes and potential generation changes within short (1h), mid (3 and 6h) and long-term (12h) timeframes, however, would be of high value for power system models.

As real generation time series are necessary for bias-correction or validation anyhow, instead of using power-curve based models, machine learning (ML) models can be applied instead to derive synthetic time series from climate data for time periods where no observed generation is available.

In particular, neural networks are a promising approach here. They can fit arbitrary, non-linear functions as they are universal function approximators (Cybenko, 1989). The need for a correction of systematic biases as a separate work step and the need for information on the accurate turbine locations and other specifications can therefore possibly be overcome when using machine learning wind power generation models. This decreases the effort when generating time series of wind power electricity generation, as gathering accurate information on turbine locations can be time-consuming or even impossible for some countries. Additionally, while machine learning (ML) models based on the same underlying climatic data cannot be expected to fully solve the problem of the correct representation of real wind power variability, they may nevertheless increase quality of results in terms of extreme values by learning spatio-temporal relationships between climatic input and (extreme) values on the output.

We consequently assess, if machine learning (ML) models are equally or even better suited than Renewables.ninja (RN), a cutting-edge power curve based model time series (Staffell and Pfenninger, 2016) to replicate the distributional, seasonal, extreme value and power ramp characteristics of actual wind power generation. Furthermore, we quantify how the quality of the ML modelled time series depends on the extent of information on wind turbine locations the ML model receives. The successful application of ML models for short and medium-term predictions of wind power (Chang et al., 2017; Heinermann and Kramer, 2016; Khosravi et al., 2018; Treiber et al., 2016) provide reasonable arguments to use MLMs for the purpose of generating synthetic wind power generation time series. However, MLMs have only been utilised to derive multi-year wind power generation time series or for predictions on spatial dimensions larger than single power plant sites or wind farms within close spatial proximity (Aghajani et al., 2016; Dong et al., 2017; Nourani Esfetang and Kazemzadeh, 2018; Park and Park, 2019). Countrywide estimation of wind power time series by means of ML models for use in energy system models has not been used before to the best of our knowledge.

We therefore apply a multilayer perceptron neural network to climate variables from a reanalysis dataset in order to generate two MLM derived time series of nationally aggregated wind power generation for Germany in capacity factors (CF). The two models differ in terms of whether information on turbine locations is made available to them. Consequently, the two resulting time series are compared with time series generated by RN in terms of model error metrics and of the representation of distributional, seasonal, extreme events and power ramp characteristics for the period of 2012-2016. We thus aim at understanding to what extent ML models can replace power-curve based models.

## Methods and Data

Neural networks are commonly cited algorithms for short and mid-term prediction of wind power generation (Abhinav et al., 2017; Chang et al., 2017; Díaz et al., 2018; Foley et al., 2012). Neural networks can be used to approximate universal functional relationships between input and output variables. In our case, the input variables consist of climate variables and additional dummies, while the output is the wind power generation time series. We choose Germany as a modelling location due to its highly developed wind turbine fleet, in addition to the sound data availability via the Open Power System Data (OPSD) platform. The model time period (2010 – 2016) has been chosen to fit the longest openly available coherent time series of hourly resolved wind power generation needed for model training. The temporal resolution of observed generation as well as the modelled generations is hourly. The input dataset for the time series MLM1 does not feature any location information at all, whereas the input dataset for MLM2 contains information on turbine locations via climate variable subsetting, which corresponds to using only the 4 nearest grid points to locations where wind turbines are actually installed. A third time series (MLM3) based on an input dataset with a more substantial subsetting of grid points has also been tested and is included in the Appendix.

### Model setup and training

We use a neural network with one input layer with a node size equal to the number of input predictor variables, i.e. our dummy variables and wind speed components, three hidden layers of a user-defined size, i.e. we tested 60 and 80 nodes and one output layer of size one, i.e. electricity generation from wind power which is the predicted variable. Out of these two network sizes the resulting time series with the better correlation, normalised root mean square error (NRMSE) and normalised mean absolute error (NMAE) values is used

In order to reduce the number of model training iterations needed to generate the seven years time series, as well as to provide a sufficient time period for model training and to comply to the training/prediction ratios' rule of thumb (2/3 of the dataset used for training; 1/3 used for prediction), the training time period needs to be rearranged for every prediction period. For the prediction period of 2010 and 2011, the training period is set to 2012 – 2016. For the prediction period 2012 and 2013, the years 2010 – 2011 and 2014 – 2016 are used for training, and similarly for all other two-year periods. This results in a time series split between training and prediction period amounting to approximately 71% to 29%, (Figure 1).

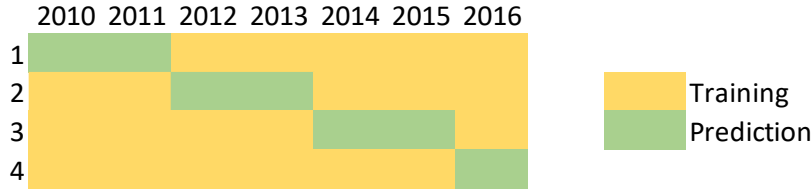

Figure 1: Network training and prediction timeframes

Preparatory steps for the MLM approach consist of acquiring the necessary climate input data and deriving date dummy variables for MLM1. An additional subsetting of climate data grid points is needed for the MLM2 dataset. Subsequently the input data are used in the model set-up steps, where the actual conversion from input data to power generation takes place. The model set-up can be split into the two tasks of model training and prediction (using the previously trained model). In the first step the neural network is trained using the input data from the training period (training dataset), after setting the seed for the random number generator to guarantee reproducibility of results. The trained neural network is consequently fed with the remaining set of input variables (prediction dataset) to compute the modelled electricity generation from wind power for the prediction period in terms of capacity factors (Figure 2).

We compare our results to the RN Data set. A detailed description of the modelling approach used by RN can be found in Staffell and Pfenninger (2016). Figure 2 shows a brief comparison of our modelling approach and RN. In contrast to the rather simple ML modelling process, RN has to acquire wind speed data, interpolate wind speeds to the turbine locations, extrapolate wind speeds to the corresponding turbine hub heights and execute an additional bias correction step for calibrating results against observations in addition to the data gathering and conversion steps (Figure 2).

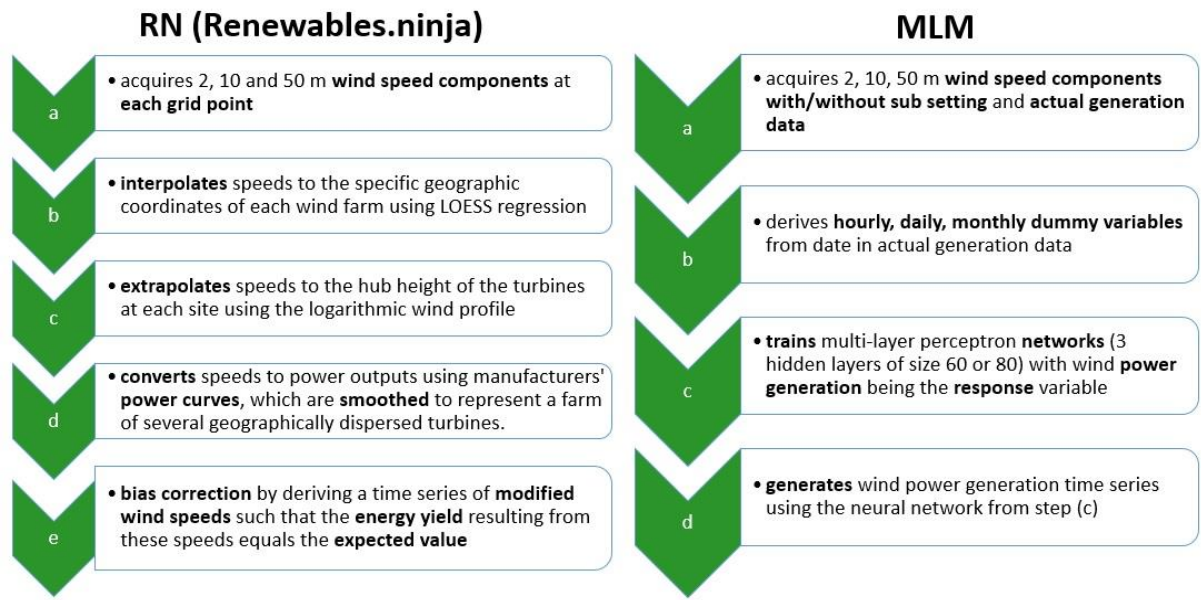

Figure 2: Modelling flow diagram (Renewables.ninja abbreviated as RN, machine learning model abbreviated as MLM, own figure)

Function calls and computations within the neural network are executed according to the specifications in Bergmeir and Benítez (2012) and the package documentation of the R-Package "caret" (Kuhn, 2018). The model training part consists of a call to a training function, that outputs a neural network model. This model is then used to derive predictions with the prediction part of the time series.

All computations and visualisations are done in RStudio (Version 1.1.423) with R version named "Microsoft R Open 3.4.3". Packages "tidyverse", "lubridate", "ggplot2" and their corresponding dependencies are used for data handling and visualisation. For model set up, training, and predictions the package "caret" is used, which itself depends on the "RSNNS" package. Downloads and handling of MERRA-files are done with the R-package called "MERRAbin" and its dependencies (joph, 2019). The source code for this methodological approach and the validation is available in a GitHub repository (jbaumg, 2019).

Climate Input Data

The present study is based on climate input variables from the global reanalysis data set MERRA2. This dataset has been chosen to enable comparison of our results to the time series from RN (ninja_europe_wind_v1.1-data package), which uses the same data source. The climate input data used in the MLM are featured in the time averaged single level diagnostics subset "tavg1_2d_slv_Nx", whereby wind speed components U2M, V2M, U10M, V10M, U50M, and V50M, i.e. wind speeds at 2 meter, 10 and 50 meter above ground were used.

Additionally, hourly, daily (weekdays), and monthly dummy variables were added for the purpose of approximating diurnal, weekly and seasonal patterns. These variables for all grid points within a bounding box (longitude from 5 to 15.625, latitude from 46 to 56) around Germany constitute the input data set for the first MLM generated time series (MLM1), where no variable sub setting is performed. For the second MLM generated time series (MLM2), the climate variable grid points are selected according to turbine locations, whereby the four nearest grid pints to actual wind turbine locations are used.

Installed Capacity and Electricity Generation Data

Observed electricity generation from OPSD ("Data Platform – Open Power System Data," 2018a) were used as the response variable for the training of the neural network for both MLM time series as well as for assessing the quality of the modelled electricity generation time series.

These generation values are consequently converted to capacity factors (CF) by dividing them by daily values of installed wind power capacity from OPSD ("Data Platform – Open Power System Data," 2018b). Locations of wind turbines taken from OPSD ("Data Platform – Open Power System Data," 2017) are additionally used to spatially subset climate data by extracting values of the four grid points nearest to wind farm locations for MLM2. The installed capacity time series does not explicitly feed into any of the two MLMs for reasons of comparability (RN time series are based on a single installed capacity value). The 30-minute time shift of the climate input data compared with observed wind power generation data used for model training and validation was neglected as the climate input data are hourly time averaged variables. Consequently, all variables were scaled by subtracting the mean and divided by the value range (minimum subtracted from maximum value). Feature scaling like this is the standard procedure in order to reduce computational effort in training of neural networks. A list of all variables used in this study, their application, unit, size and source can be found Table 1.

Table 1: All variables used in the MLM modelling process and validation with their corresponding name, application, unit, size in MLM1, size in MLM2 and source (in the size column the multiplicator corresponds to the number of grid points used and the multiplicand corresponds to the time series length in hours from 2010 to 2016; 43 date dummy variables: 24 hour, 7 day and 12 month dummy variables)

| Name | Application | Unit | MLM1 Size | MLM2 Size | Source |
|---|---|---|---|---|---|
| 2m northward wind speed (V2M) | predictor | m/s | 378x61368 | 206x61368 | MERRA2 |
| 10m northward wind speed (V10M) | predictor | m/s | 378x61368 | 206x61368 | MERRA2 |
| 50m northward wind speed (V50M) | predictor | m/s | 378x61368 | 206x61368 | MERRA2 |
| 2m eastward wind speed (U2M) | predictor | m/s | 378x61368 | 206x61368 | MERRA2 |
| 10m eastward wind speed (U10M) | predictor | m/s | 378x61368 | 206x61368 | MERRA2 |
| 50m eastward wind speed (U50M) | predictor | m/s | 378x61368 | 206x61368 | MERRA2 |
| date dummy variables | predictor | - | 43x61368 | 43x61368 | OPSD date sequence |
| actual generation | response | capacity factor | 61368 | 61368 | OPSD |
| RN generation | validation | capacity factor | 61368 | 61368 | Renewables.ninja |
| MLM1 generation | validation | capacity factor | 61368 | 61368 | Neural network prediction |
| MLM2 generation | validation | capacity factor | 61368 | 61368 | Neural network prediction |

## Time Series Quality Assessment

Standard model error metrics (correlations, model error) and the modelled time series' ability to replicate characteristics of the observed time series (distributions, seasonal deviations, representation of extreme values and power ramps) are assessed. The comparison is based on an hourly time series covering seven generation years (2010 – 2016). Time series quality is assessed for the whole seven-year time series. However, time series quality may vary for both MLMs between different years due to the different training periods. For example, when the prediction period is set to 2012 and 2013, the years 2010 – 2011 and 2014 – 2016 are used for training, and due to possibly different turbine locations between 2011 and 2013 the training of the neural network may be less precise. We emphasize here that we always compare the RN model to time series predicted from different trained models, i.e. we do not use training periods for comparison.

# Results

## Model selection

For hyperparameter optimization, we tested two different network sizes, one with 60 and one with 80 nodes in the hidden layers. Both network sizes were used to generate the whole seven year prediction time series with all considered models. The network size featuring better model error metrics was chosen for a more thorough assessment of time series quality. When comparing error metrics for the two models MLM1 and MLM2 with a neural network of three hidden layers with 60 and with 80 nodes each for the prediction period, MLM1 performs better with a smaller network size. MLM2 performs remarkably better when using the prediction dataset with the bigger network size. We have also tested MLM3 (Appendix), which performs better with the smaller network size (Table 2).

Table 2: Comparison of model error values for all tested models for the training and the prediction dataset

| Error Metric | Network Size | MLM1 | MLM2 | MLM3 |
|---|---|---|---|---|
| NMAE | 60 | 0.152 | 0.144 | 0.138 |
| NRMSE | 60 | 0.208 | 0.202 | 0.194 |
| NMAE | 80 | 0.161 | 0.138 | 0.141 |
| NRMSE | 80 | 0.220 | 0.190 | 0.194 |

## Basic time series quality

Both MLM time series feature comparable or better error metrics than RN. MLM2 also exhibitis a similiarly high correlation as RN.

The MLM1 hourly time series features a comparable NMAE (0.152) as well as a slightly lower NRMSE (0.209) value with a slightly lower hourly correlation (COR: 0.970) compared to the RN time series (NMAE: 0.152, NRMSE: 0.210, COR: 0.976).

In comparison with the RN time series, the MLM2 hourly time series features a generally lower NMAE (0.138), and a lower NRMSE (0.191) value with a comparably high hourly correlation (COR: 0.975).

Both, MLM1 (VAR: 0.025) and MLM2 (VAR: 0.025), estimate the total observed time series' variance (VAR: 0.026) more accurately compared to RN (VAR: 0.030). When quantiles are considered, both MLMs are closer to the observed quantiles than RN except for the 0% quantile where the MLMs are lower due to presence of negative values and except for the 25% and the 100% quantile for the MLM1 and the 100% quantile for the MLM2 time series. The MLM1 time series features 53 negative capacity factor events and the MLM2 65 thereof (Table 3).

Table 3: Comparison of error metrics for RN, MLM1 and MLM2 approaches

| Quality Measure | Observations | RN | MLM1 | MLM2 |
|---|---|---|---|---|
| Correlation | - | 0.976 | 0.970 | 0.975 |
| NMAE | - | 0.152 | 0.152 | 0.138 |
| NRMSE | - | 0.210 | 0.209 | 0.191 |
| Variance | 0.026 | 0.030 | 0.025 | 0.025 |
| Quantile | | | | |
| 0% | 0.000 | 0.000 | -0.012 | -0.011 |
| 25% | 0.067 | 0.071 | 0.072 | 0.069 |
| 50% | 0.137 | 0.149 | 0.140 | 0.135 |
| 75% | 0.260 | 0.275 | 0.260 | 0.254 |
| 100% | 0.913 | 0.964 | 0.970 | 0.975 |

Except for the highest capacity factor (CF) range, both MLM time series mainly fare equally well or better than the RN when deviations from observed values are considered.

Median deviations in five CF ranges (except for the CF ranges from 0.0-0.1, 0.4-0.5, 0.5-0.6, and above 0.8) are less distant to zero when comparing with the MLM1 time series. With the MLM2 time series also the median deviation for five CF ranges is less distant to zero than with the RN time series (except for the CF ranges from 0.0-0.1, 0.3-0.4, 0.4-0.5 and above 0.8). With the RN time series an overestimation is apparent in the scatterplot, particularly in the higher CF spectrum. The ranges of deviations for MLM1 are slightly wider than with the RN time series except for the values between 0.0-0.1, 0.1-0.2 and 0.2-0.3. In a similar way, the deviation ranges of MLM2 are only narrower than the RN deviations in two CF classes (CF 0.0-0.1, CF 0.1-0.2). The deviation median in the MLM2 time series is generally less distant to zero than in the MLM1 time series except for the CF classes from 0.2-0.3 and 0.3-0.4, for the deviation ranges the picture is fairly indifferent with the MLM2 time series featuring a narrower range in five CF ranges (0.0-0.1; 0.4-0.5; 0.5-0.6; 0.6-0.7; 0.7-0.8). This is also reflected in the scatterplot with the MLM2 values more concentrated towards the diagonal line than the MLM1 values. Both MLM time series also mainly underestimate generation slightly within most CF classes, whereas the RN time series overestimates (Figure 3).

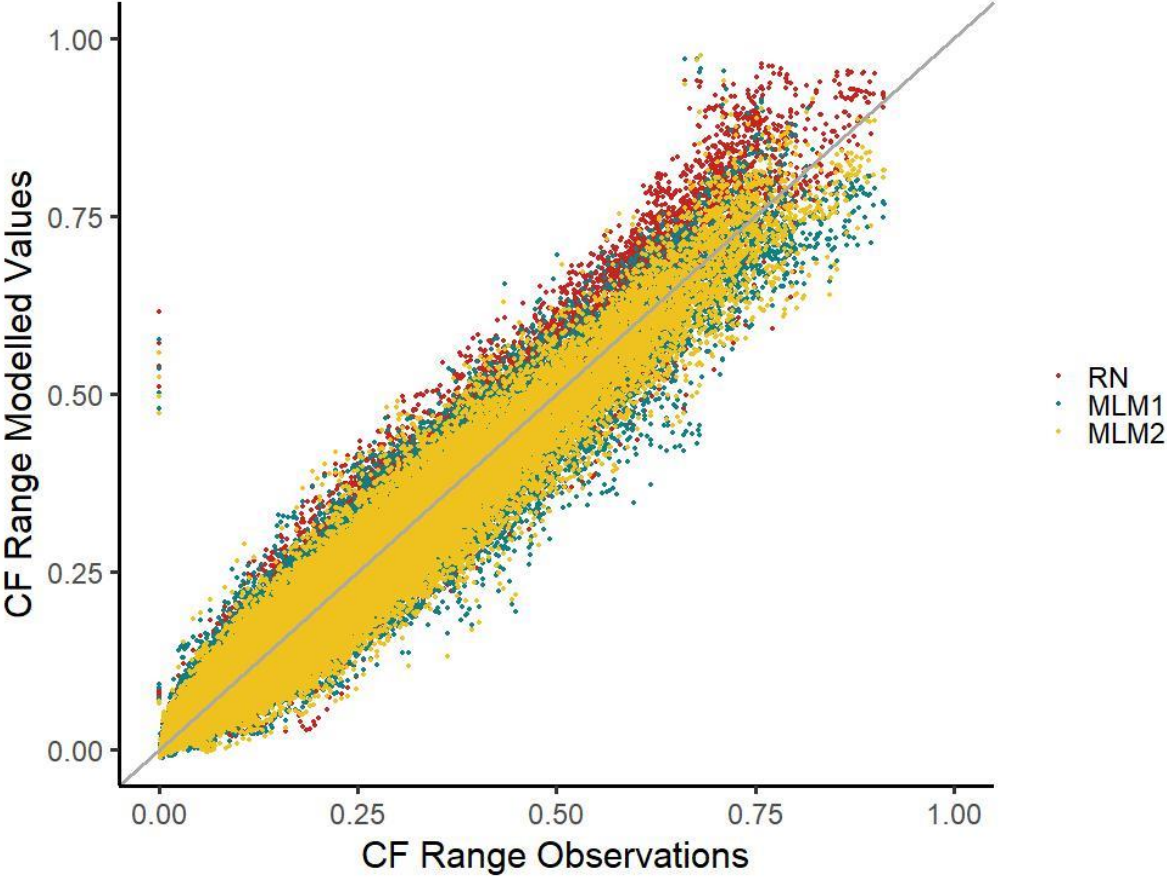

Figure 3: Scatterplot of modelled generation time series compared with observations (Renewables.ninja time series abbreviated as RN and machine learning model time series abbreviated as MLM1 and MLM2 versus observations on the x-axis)

The MLM1 and MLM2 time series reduce the underestimation of CF values around 0.1 visible in the RN time series. However, both MLMs underestimate occurrences of CFs in the very low range compared with RN.

The results show an acceptable representation of the distributional characteristics of the observed time series comparable to the results of RN for both MLM time series. Generation in the very low CF range (8,254 actual values with a CF < 0.04) are better approximated by the RN time series (8,138 modelled values with a CF < 0.04) than by MLM1 (6,658 modelled values) or MLM2 (7,115 modelled values), which is also represented by a higher density in the RN time series. CFs in the range around 0.1 (8,928 actual values between 0.08 and 0.12), where an overestimation occurs in the MLM-derived time series (9,146 modelled values for the MLM1 time series and 9,249 for MLM2) opposed to an underestimation in the RN time series (7,989 modelled values), however are better represented by the MLMs. This can also be seen by a better fit of the probability density curve in this range for both MLMs. In the high CF range, the frequency of CF values above 0.8 are, although slightly underestimated (75 and 126 values > 0.8 for the MLM1 and MLM2 versus 121 actual values), better approximated by MLM time series than by RN (471 modelled values > 0.8) (Figure 4).

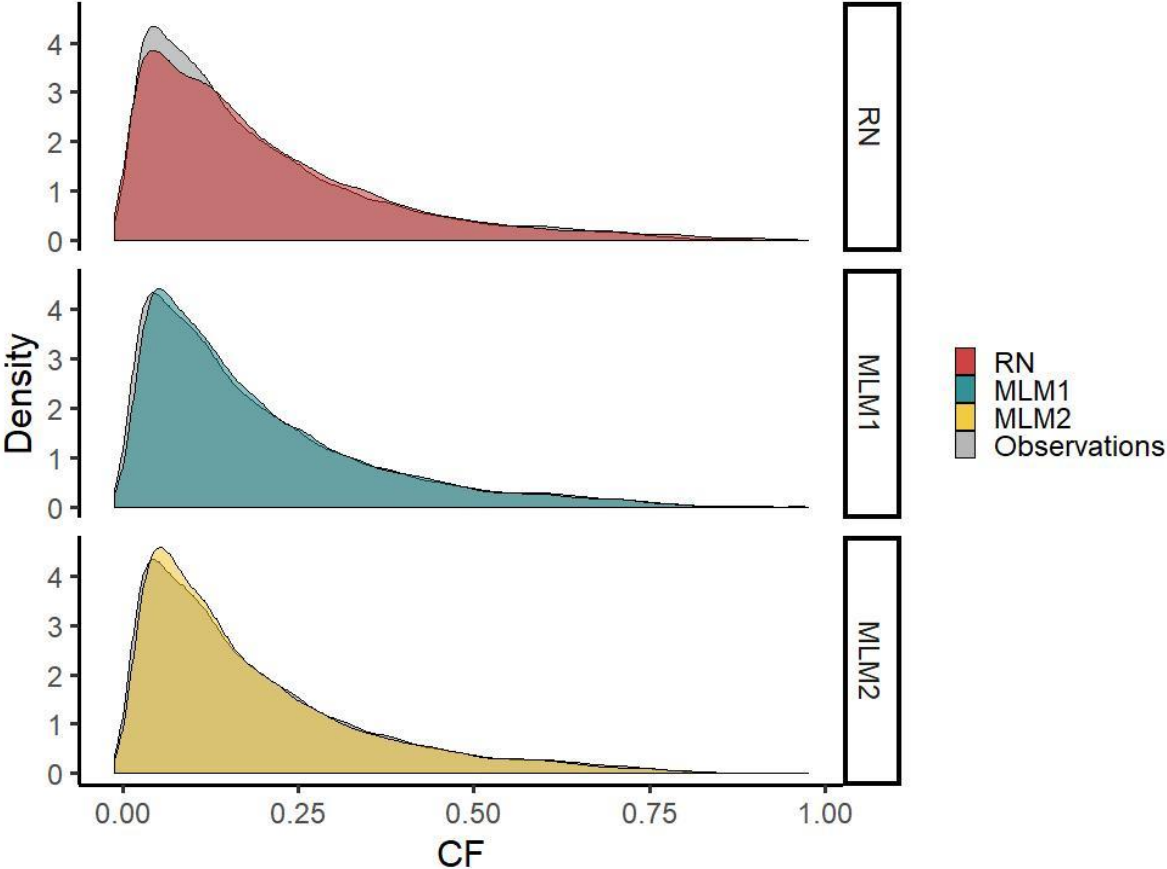

Figure 4: Total distributions of the RN, MLM1 and MLM2 modelled time series compared with observed total distribution (dark coloured areas correspond to an accordance of the modelled and the observed time series; Renewables.ninja time series abbreviated as RN and machine learning model time series abbreviated as MLM1 and MLM2)

### Diurnal Characteristics

MLM2 provides a better estimate of diurnal characteristics for most hours and both MLM time series reduce deviations in the evening hours featured in the RN time series (*Figure 5*).

RN median deviations are generally lower than with MLM1 except for seven hours (hours 2, 15, 16, 17, 18, 19, 20). For the MLM2 the median deviations are generally lower than with the RN time series except for four hours (hours 4, 6, 22, 23). For RN the mean of deviations is highest around the evening (hours 18, 19, 20), where it is noticeably higher than during the rest of the day. It is lowest during morning (hours 7, 8) and around midnight (hours 1, 24), where it is lower than during the remaining hours. The MLM1 and MLM2 time series do not feature a similarly strong increase of the deviation median in the evening hours as it is the case with the RN time series.

The deviation range of the MLM1 time series is narrower for 8 hours (hours 1, 2, 3, 4, 5, 6, 22, 23) compared with the RN time series and for 12 hours (1, 2, 3, 4, 5, 6, 7, 9, 21, 22, 23, 24) when the MLM2 time series is compared. For all three time series the deviation range is remarkably wider in the early morning hours 2 – 5 than during the remaining hours. With all three time series (most remarkably with the MLM1 time series) an increase of the deviation range around noon and early afternoon can be seen and for all three time series the deviation range decreases again for the evening hours (Figure 5).

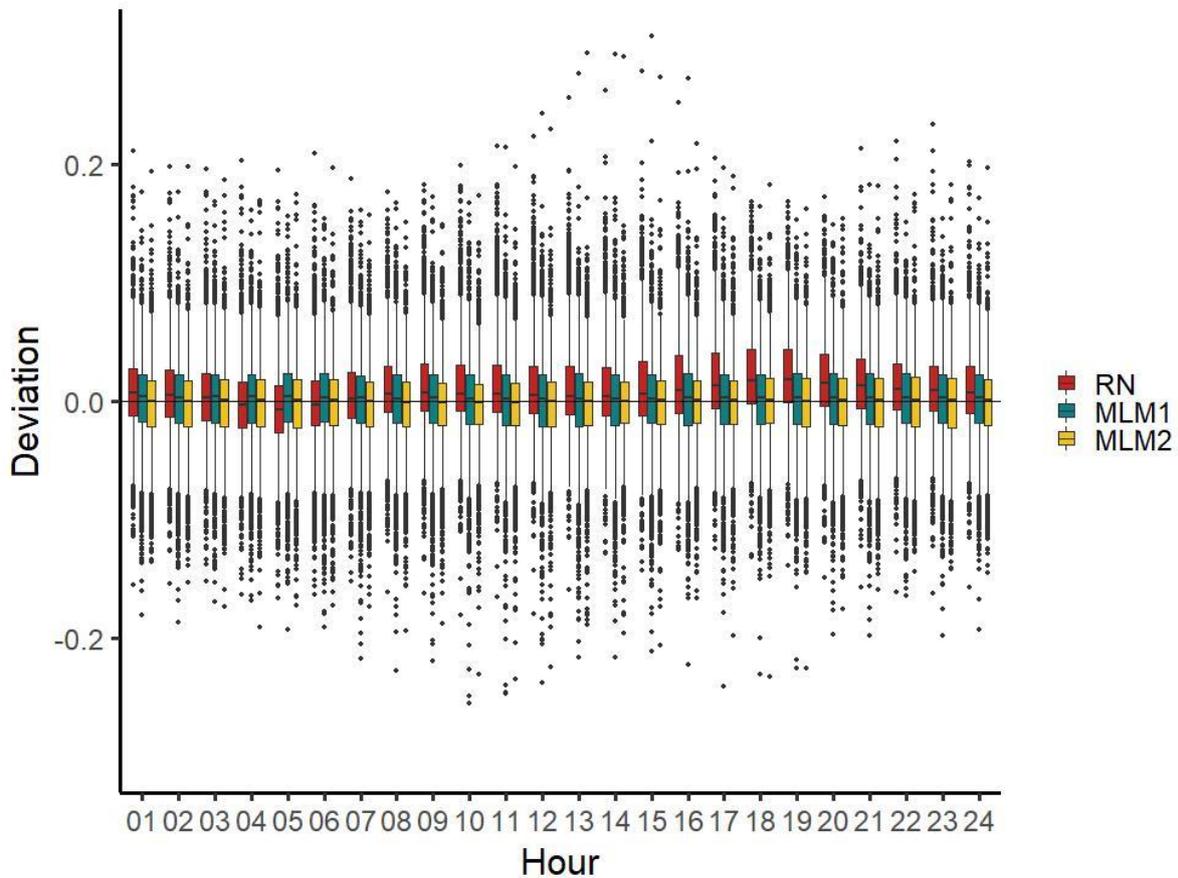

Figure 5: Hourly deviations from observed values for RN, MLM1 and MLM2 (Renewables.ninja time series abbreviated as RN and machine learning model time series abbreviated as MLM1 and MLM2)

Seasonal characteristics

All three time series reflect seasonal time series characteristics comparably well. However, MLM2 performs better than the RN time series in the majority of CF classes and seasons. Interestingly, the lowest CF class in the winter season features four outliers in all three time series.

When comparing observations with the MLM1 time series the median deviation is lower than for RN for 9 out of 19 classes and seasons, where the MLM1 is better in the CF classes 0.2-0.4 and 0.6-0.8. The MLM2 time series performs better in 11 out of 19 classes and seasons with CF classes 0-0.2 and 0.6-0.8 being the sweet spot. For all three time series, a tendency to overestimate values in the lowest CF class can be observed for all seasons. In the remaining

CF classes and seasons, the MLM time series mainly underestimate, where the RN time series overestimates. For summer no observed values in the CF class >0.8 are featured. Upon comparison of the deviation ranges (spread from minimum to maximum deviation value), MLM2 fares better than MLM1 outperforming the RN time series in 9 out of 19 classes and seasons. MLM1 provides a fairly indifferent picture with respect to deviation ranges, however for most classes and seasons it performs worse than RN. Remarkably, the lowest CF class in the winter season features some outliers skewing all three time series towards a high deviation range for this class and season. For the deviation mean as well as the range of deviations, MLM2 outperforms RN in more CF classes than MLM1 (Figure 6).

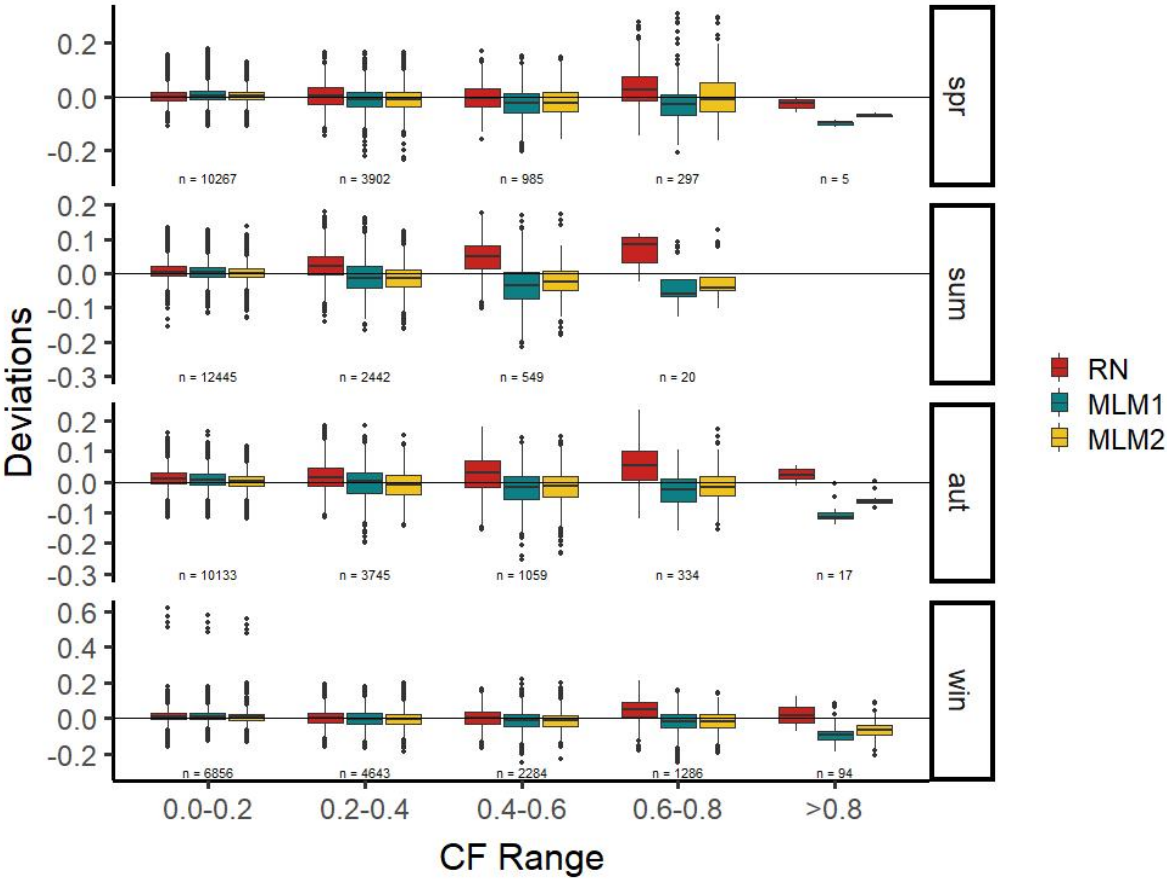

Figure 6: CF deviations of RN, MLM1 and MLM2 modelled time series within different CF bins for separate seasons from 2010 to 2016 (Renewables.ninja time series abbreviated as RN and machine learning model time series abbreviated as MLM1 and MLM2, n denotes the number of observations within capacity factor bins)

## Durations and frequencies of low, high and extreme values

> MLM1 and MLM2 provide a better estimate of frequencies and durations of low CF extremes and frequencies of high CF extremes. Durations of high CF extremes are better approximated by RN.

Frequencies (102 actual events) of low capacity factor extreme values (CF < 0.005) are better approximated by MLM1 (193 modelled events) and MLM2 (238 modelled events) than by the RN time series (430 events), where both MLMs provide a less pronounced overestimation and therefore better approximation. Mean durations (3.19 consecutive hours of observed generation below 0.005 CF) of low capacity factor extreme values are also better approximated by MLM1 (3.27 consecutive hours of observed generation below 0.005 CF) and MLM2 (3.50 consecutive hours of observed generation below 0.005 CF) compared with RN (4.62 consecutive hours of modelled generation below 0.005 CF). Both MLM time series provide an exact match of low generation extremes maximum duration (10 observed consecutive hours below 0.005 CF), where the RN time series overestimates maximum duration (14 consecutive hours of modelled generation below 0.005 CF). Frequencies (121 actual events) of high capacity factor extreme values (CF > 0.8) are better approximated by MLM1 (75 modelled events) and MLM2 (125 modelled events) than by RN (471 modelled events). When mean durations of very high generation events (7.56 observed consecutive hours) are considered, RN (9.24 modelled consecutive hours) provides the best estimate compared with MLM1 (4.17 modelled consecutive hours) and MLM2 (5.21 modelled consecutive hours). With regard to the approximation of maximum durations (35 consecutive observed hours) the MLM1 time series (11 modelled consecutive hours) is fairly far off, where RN (33 modelled consecutive hours) and MLM2 (21 modelled consecutive hours) provide the better estimates (Table 3).

Table 3: Observed and modelled (Renewables.ninja [RN] and machine learning models [MLM1, MLM2]) CF including frequencies, mean and maximum durations of extremely low (<0.005), low (0.01), high (0.75) and extremely high (>0.8) generation events in consecutive hours

| CF Ranges | Observations | RN | MLM1 | MLM2 |
|---|---|---|---|---|
| CF <0.005 | | | | |
| Frequency | 102 | 430 | 193 | 238 |
| Mean Duration | 3.19 | 4.62 | 3.27 | 3.50 |
| Max Duration | 10 | 14 | 10 | 10 |
| CF <0.01 | | | | |
| Frequency | 509 | 748 | 312 | 341 |
| Mean Duration | 2.98 | 2.60 | 1.91 | 1.91 |
| Max Duration | 25 | 12 | 7 | 5 |
| CF >0.75 | | | | |
| Frequency | 204 | 356 | 174 | 225 |
| Mean Duration | 3.46 | 3.10 | 3.63 | 3.41 |
| Max Duration | 13 | 18 | 22 | 15 |
| CF >0.8 | | | | |
| Frequency | 121 | 471 | 75 | 125 |
| Mean Duration | 7.56 | 9.24 | 4.17 | 5.21 |
| Max Duration | 35 | 33 | 11 | 21 |

Both MLM time series follow the cumulative density of capacity factor changes within one hour better than the RN time series. In particular, the RN time series overestimates density within the range of negative CF changes between -0.1 and 0 and underestimates in the range of positive CF changes between 0 and 0.1. MLM1 and MLM2 do not show this behaviour and nearly match the observed CF change density exactly (Figure 7).

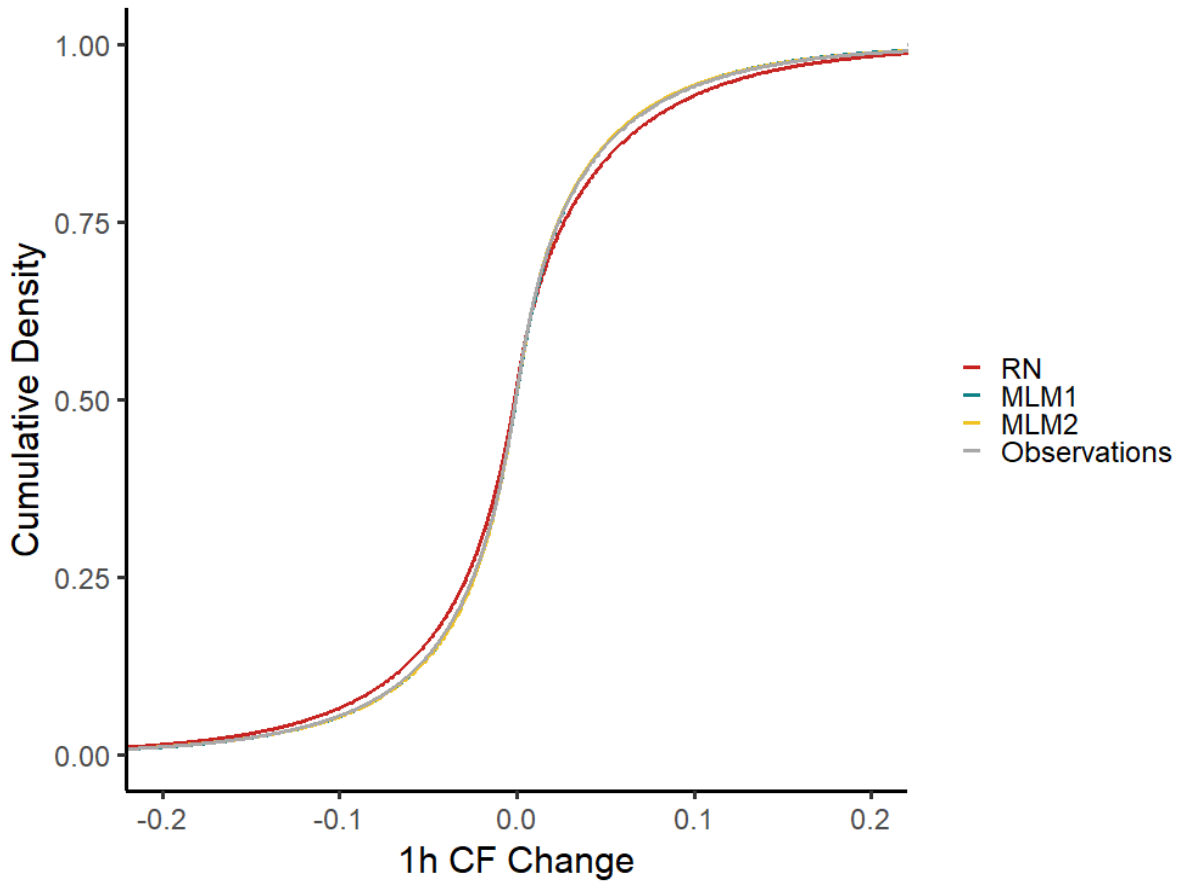

*Figure 7: Cumulative density of one hour CF changes for the RN, MLM1 and MLM2 modelled time series compared with observations*

### Frequencies and ranges of power ramps

Within all considered time frames, both MLM time series replicate CF changes better or equally well than the RN time series.CF changes within four different time frames (1h, 3h, 6h, 12h) are better or equally well replicated with MLMs compared to RN. Both MLM time series feature mean values of positive and negative CF changes, frequencies of negative and positive CF changes, minimum and maximum ramp values and an approximation of the frequencies of fairly high and low power ramps comparable to or better than the RN within nearly all time frames. All models fare better the longer the time frame, reflecting the ability of the reanalysis data source to replicate the temporal variability of wind speeds within longer time frames (Table 4).

Table 4: Minimum, Maximum, Mean of negative and positive CF ramps within different time frames and frequency of highly positive and negative CF ramps

| Time frame | Observations | RN | MLM1 | MLM2 |
|---|---|---|---|---|
| 1h | | | | |
| Min | -0.66 | -0.11 | -0.14 | -0.13 |
| Neg. Mean | -0.012 | -0.013 | -0.013 | -0.012 |
| Neg. Freq | 31313 | 32243 | 31070 | 31132 |
| Pos. Mean | 0.013 | 0.014 | 0.013 | 0.013 |
| Pos. Freq | 30054 | 29124 | 30297 | 30235 |
| Max | 0.50 | 0.16 | 0.16 | 0.14 |
| Frequency<-0.2 | 1 | 0 | 0 | 0 |
| Frequency>0.2 | 1 | 0 | 0 | 0 |
| 3h | | | | |
| Min | -0.67 | -0.28 | -0.30 | -0.33 |
| Neg. Mean | -0.022 | -0.024 | -0.023 | -0.022 |
| Neg. Freq | 94293 | 96677 | 93587 | 94091 |
| Pos. Mean | 0.023 | 0.027 | 0.023 | 0.023 |
| Pos. Freq | 89805 | 87421 | 90411 | 90007 |
| Max | 0.50 | 0.30 | 0.38 | 0.31 |
| Frequency<-0.2 | 75 | 71 | 84 | 61 |
| Frequency>0.2 | 92 | 181 | 85 | 93 |
| 6h | | | | |
| Min | -0.67 | -0.46 | -0.48 | -0.50 |
| Neg. Mean | -0.035 | -0.038 | -0.035 | -0.034 |
| Neg. Freq | 188949 | 192660 | 188102 | 188749 |
| Pos. Mean | 0.037 | 0.042 | 0.036 | 0.036 |
| Pos. Freq | 179238 | 175527 | 180085 | 179438 |
| Max | 0.51 | 0.50 | 0.55 | 0.47 |
| Frequency<-0.2 | 1,616 | 2,140 | 1,678 | 1,736 |
| Frequency>0.2 | 1,846 | 2,921 | 1,650 | 1,860 |
| 12h | | | | |
| Min | -0.68 | -0.60 | -0.66 | -0.64 |
| Neg. Mean | -0.054 | -0.059 | -0.053 | -0.053 |
| Neg. Freq | 376200 | 380753 | 375262 | 376127 |
| Pos. Mean | 0.056 | 0.064 | 0.055 | 0.055 |
| Pos. Freq | 360138 | 355585 | 361076 | 360211 |
| Max | 0.62 | 0.67 | 0.65 | 0.68 |
| Frequency<-0.2 | 14,185 | 18,009 | 13,781 | 14,245 |
| Frequency>0.2 | 14,680 | 19,194 | 13,634 | 14,237 |

## Discussion

The MLM approach can only be successfully applied if (1) sufficiently long and high quality climate input data as well as generation data are available for the prevailing wind conditions and turbine locations in a region, which currently is not the case for all regions. This however does not translate to a downside of using a neural network approach as this also holds true for power curve based models, as they are in need for observations for calibration or validation. (2) Both MLMs are not capable to reflect changes in the spatial configuration of installed wind turbine capacity as installed capacities are not used as model inputs. However, the proposed approach can be easily adapted to make use of information on installed capacities.

The proposed approach of deriving time series by means of neural networks is only (3) partly suited to generate future scenarios taking significant technological developments into account (e.g. a considerable shift of the ratio between rotor diameter and installed capacity such as in turbines specifically designed for low wind speed conditions) compared with a power curve based approach such as in RN. Using turbine specifications as an input to model training can probably compensate for this downside in subsequent model iterations, although this would require additional technical information on the turbine types used. (4) Landmark changes in technology (e.g. horizontal turbine design) or the regulation of wind turbines or intermittent renewables (e.g. increased curtailment) in general, for which no observational data are available cannot be successfully replicated by the proposed neural network approach. This issue can be addressed for both methodologies using post treatment of the model output. (5) For the MLM training step, a significant amount of computational effort is required, which is probably higher than the computational effort associated with the model setup of power curve based models. The prediction step, however, is comparable in computational complexity to power curve based approaches.

## Conclusions

Both machine learning models have been able to generate wind power generation time series comparable to or even better than a state of the art power curve based modelling approach (Renewables.ninja abbreviated as RN) with respect to standard error metrics, seasonal distributional characteristics, frequencies and durations of low, high and extreme values as well as for the replication of frequencies and durations of power ramps for wind power generation.

We used two datasets, one without location information and one with implicit location information via climate data grid point subsetting as an input to a MLM to assess whether location information is necessary to obtain time series quaity comparable to RN. We found that (1) both input datasets to the machine learning model time series have been able to generate wind power generation time series comparable to or even better than a state of the art power curve based modelling approach (RN) with respect to the quality measures considered.

(2) Both MLM generated time series (especially MLM2) additionally show reduced overestimation of very high CF values and reduced underestimation of CF values around a CF of 0.1. (3) Time series quality varies depending on which quality parameter is considered and on which MLM is used. However, there are no major drawbacks of using a machine learning approach for the purpose of generating wind power time series with the exception of the duration of high generation extreme events, which are however rare.

(4) The presence of negative CF values indicates that longer training periods could be helpful as the whole range of all input variable combinations and thus the variability of climate data has not been fully captured by the available training time frames. However, negative values can easily be handled via simple post-processing. The MLM approach can be seen as a valuable addition to traditional power curve based modelling concepts given long enough time series of actual generation for model training. (5) Furthermore, the information required for model setup with regards to knowing accurate wind turbine locations and power curves is much lower. The additional information on turbine locations and the used turbine models is not strictly needed to reach a time series quality comparable to the RN approach, although implicit location information improves most of the time series quality measures considered as MLM2 outperforms MLM1 in most cases.

**Acknowledgements**

We acknowledge funding by CLIM2POWER. Project CLIM2POWER is part of ERA4CS, an ERA-NET initiated by JPI Climate, and funded by FORMAS (SE), BMBF (DE), BMWFW (AT), FCT (PT), EPA (IE), ANR (FR) with co-funding by the European Union (Grant 690462). We also gratefully acknowledge support from the European Research Council ("reFUEL" ERC-2017-STG 758149) and would like to cordially thank Iain Staffell for the provision of historic observed generation data used for Renewables.ninja model calibration.

# Bibliography

Abhinav, R., Pindoriya, N.M., Wu, J., Long, C., 2017. Short-term wind power forecasting using wavelet-based neural network. Energy Procedia 142, 455–460. https://doi.org/10.1016/j.egypro.2017.12.071

Aghajani, A., Kazemzadeh, R., Ebrahimi, A., 2016. A novel hybrid approach for predicting wind farm power production based on wavelet transform, hybrid neural networks and imperialist competitive algorithm. Energy Conversion and Management 121, 232–240. https://doi.org/10.1016/j.enconman.2016.05.024

Andresen, G.B., Søndergaard, A.A., Greiner, M., 2015. Validation of Danish wind time series from a new global renewable energy atlas for energy system analysis. Energy 93, 1074–1088. https://doi.org/10.1016/j.energy.2015.09.071

Arbeitsgemeinschaft Energiebilanzen, 2019. Bruttostromerzeugung in Deutschland ab 1990 nach Energieträgern.

Bergmeir, C., Benítez, J.M., 2012. Neural Networks in *R* Using the Stuttgart Neural Network Simulator: **RSNNS**. Journal of Statistical Software 46. https://doi.org/10.18637/jss.v046.i07

Bonjean Stanton, M.C., Dessai, S., Paavola, J., 2016. A systematic review of the impacts of climate variability and change on electricity systems in Europe. Energy 109, 1148–1159. https://doi.org/10.1016/j.energy.2016.05.015

Cannon, D.J., Brayshaw, D.J., Methven, J., Coker, P.J., Lenaghan, D., 2015. Using reanalysis data to quantify extreme wind power generation statistics: A 33 year case study in Great Britain. Renewable Energy 75, 767–778. https://doi.org/10.1016/j.renene.2014.10.024

Chang, G.W., Lu, H.J., Chang, Y.R., Lee, Y.D., 2017. An improved neural network-based approach for short-term wind speed and power forecast. Renewable Energy 105, 301–311. https://doi.org/10.1016/j.renene.2016.12.071

Cybenko, G., 1989. Approximation by superpositions of a sigmoidal function 12.

Data Platform – Open Power System Data [WWW Document], 2018a. URL https://data.open-power-system-data.org/time_series/ (accessed 6.28.18).

Data Platform – Open Power System Data [WWW Document], 2018b. URL https://data.open-power-system-data.org/renewable_power_plants/ (accessed 8.16.18).

Data Platform – Open Power System Data [WWW Document], 2017. URL https://data.open-power-system-data.org/renewable_power_plants/2017-07-03/ (accessed 8.17.18).

Díaz, S., Carta, J.A., Matías, J.M., 2018. Performance assessment of five MCP models proposed for the estimation of long-term wind turbine power outputs at a target site using three machine learning techniques. Applied Energy 209, 455–477. https://doi.org/10.1016/j.apenergy.2017.11.007

Dong, Q., Sun, Y., Li, P., 2017. A novel forecasting model based on a hybrid processing strategy and an optimized local linear fuzzy neural network to make wind power forecasting: A case study of wind farms in China. Renewable Energy 102, 241–257. https://doi.org/10.1016/j.renene.2016.10.030

E3M, 2018. Manual of PRIMES [WWW Document]. URL http://www.e3mlab.eu/e3mlab/index.php?option=com_content&view=article&id=58%3Amanual-for-primes-model&catid=35%3Aprimes&Itemid=80&lang=en (accessed 9.1.19).

Foley, A.M., Leahy, P.G., Marvuglia, A., McKeogh, E.J., 2012. Current methods and advances in forecasting of wind power generation. Renewable Energy 37, 1–8. https://doi.org/10.1016/j.renene.2011.05.033

Global Wind Energy Council, 2019. Global Wind Report 2018.

Global Wind Energy Council, 2008. GLOBAL WIND 2007 REPORT.

Gruber, K., Klöckl, C., Regner, P., Baumgartner, J., Schmidt, J., 2019. Assessing the Global Wind Atlas and local measurements for bias correction of wind power generation simulated from MERRA-2 in Brazil. Energy 116212. https://doi.org/10.1016/j.energy.2019.116212

Heinermann, J., Kramer, O., 2016. Machine learning ensembles for wind power prediction. Renewable Energy 89, 671–679. https://doi.org/10.1016/j.renene.2015.11.073

jbaumg, 2019. jbaumg/Machine-learning-long-term-wind-power-time-series.

joph, 2019. joph/MERRAbin.

Khosravi, A., Koury, R.N.N., Machado, L., Pabon, J.J.G., 2018. Prediction of wind speed and wind direction using artificial neural network, support vector regression and adaptive neuro-fuzzy inference system. Sustainable Energy Technologies and Assessments 25, 146–160. https://doi.org/10.1016/j.seta.2018.01.001

Klein, D.R., Olonscheck, M., Walther, C., Kropp, J.P., 2013. Susceptibility of the European electricity sector to climate change. Energy 59, 183–193. https://doi.org/10.1016/j.energy.2013.06.048

Kuhn, M., 2018. The caret Package.

Loulou, R., Goldstein, G., Kanudia, A., Lettila, A., Remme, U., 2016. Documentation_for_the_TIMES_Model-Part-I.

Nourani Esfetang, N., Kazemzadeh, R., 2018. A novel hybrid technique for prediction of electric power generation in wind farms based on WIPSO, neural network and wavelet transform. Energy 149, 662–674. https://doi.org/10.1016/j.energy.2018.02.076

Olauson, J., Bergkvist, M., 2015. Modelling the Swedish wind power production using MERRA reanalysis data. Renewable Energy 76, 717–725. https://doi.org/10.1016/j.renene.2014.11.085

Park, Junyoung, Park, Jinkyoo, 2019. Physics-induced graph neural network: An application to wind-farm power estimation. Energy 187, 115883. https://doi.org/10.1016/j.energy.2019.115883

Réseau de transport d'électricité, 2018. Shedding light on the future of the energy system [WWW Document]. URL https://antares-simulator.org/ (accessed 9.1.19).

Simoes, S., Zeyringer, M., Mayr, D., Huld, T., Nijs, W., Schmidt, J., 2017. Impact of different levels of geographical disaggregation of wind and PV electricity generation in large energy system models: A case study for Austria. Renewable Energy 105, 183–198. https://doi.org/10.1016/j.renene.2016.12.020

Staffell, I., Green, R., 2014. How does wind farm performance decline with age? Renewable Energy 66, 775–786. https://doi.org/10.1016/j.renene.2013.10.041

Staffell, I., Pfenninger, S., 2016. Using bias-corrected reanalysis to simulate current and future wind power output. Energy 114, 1224–1239. https://doi.org/10.1016/j.energy.2016.08.068

Strbac, G., Aunedi, M., Papadaskalopoulos, D., Qadrdan, M., Pudjianto, D., Djapic, P., Konstantelos, I., Teng, F., n.d. Modelling of Smart Low-Carbon Energy Systems 7.

Tobin, I., Greuell, W., Jerez, S., Ludwig, F., Vautard, R., van Vliet, M.T.H., Bréon, F.-M., 2018. Vulnerabilities and resilience of European power generation to 1.5 °C, 2 °C and 3 °C warming. Environ. Res. Lett. 13, 044024. https://doi.org/10.1088/1748-9326/aab211

Treiber, N.A., Heinermann, J., Kramer, O., 2016. Wind Power Prediction with Machine Learning, in: Lässig, J., Kersting, K., Morik, K. (Eds.), Computational Sustainability. Springer International Publishing, Cham, pp. 13–29. https://doi.org/10.1007/978-3-319-31858-5_2

WindEurope, 2019. Wind energy in Europe in 2018 Trends and statistics.

Zucker, A., Gonzalez Aparicio, I., Careri, F., Monforti, F., Huld, T., Badger, J., European Commission, Joint Research Centre, 2016. EMHIRES dataset part I, wind power generation: European Meteorological derived HIgh resolution RES generation time series for present and future scenarios. Publications Office, Luxembourg.

# Appendix

### Results of a model with a stronger subsetting of the climate data grid points (MLM3)

For this model, we reduced the number of climate data grid points to sites with an installed capacity above the third quartile, which means only grid points with an installed capacity bigger than 75% of all grid points installed capacities are taken (58 grid points used versus 206 used in MLM2). This is only possible when not only wind farm locations are known, but also their installed capacities, which translates to a higher degree of location information that is necessary.

### Basic time series quality of MLM3

The MLM3 hourly time series features a similarly high correlation (0.975), a similarly low NMAE (0.138) and only a slightly higher NRMSE (0.194) value compared with the MLM2 time series (NMAE: 0.138, NRMSE: 0.194, COR: 0.975).

The total observed time series' variance (VAR: 0.026) is estimated only slightly less accurate with MLM3 (0.024) than with MLM2 (VAR: 0.025). When quantiles are considered, the MLM3 time series performs better for 3 quartiles (0%, 50%, 100%) when compared to MLM2. The MLM3 time series exhibits 81 negative capacity factor events compared with 65 thereof for MLM2 (Table 5).

Table 5: Comparison of error metrics for RN, MLM1 and MLM2 approaches

| Quality Measure | Observations | RN | MLM2 | MLM3 |
| --- | --- | --- | --- | --- |
| Correlation | - | 0.976 | 0.975 | 0.975 |
| NMAE | - | 0.152 | 0.138 | 0.138 |
| NRMSE | - | 0.210 | 0.191 | 0.194 |
| Variance | 0.026 | 0.030 | 0.025 | 0.024 |
| Quantile | | | | |
| 0% | 0.000 | 0.000 | -0.011 | -0.008 |
| 25% | 0.067 | 0.071 | 0.069 | 0.070 |
| 50% | 0.137 | 0.149 | 0.135 | 0.136 |
| 75% | 0.260 | 0.275 | 0.254 | 0.251 |
| 100% | 0.913 | 0.964 | 0.975 | 0.943 |

Median deviations in two capacity factor (CF) ranges (0.1-0.2, 0.6-0.7) are less distant to zero when comparing with MLM2 time series, which translates to an average underestimation stronger than with MLM2. The ranges of deviations for MLM3 are slightly narrower than with MLM2 time series for values between 0.0-0.1, 0.3-0.4, 0.4-0.5 and >0.8. Both time series look fairly similar in a scatterplot (Figure 8).

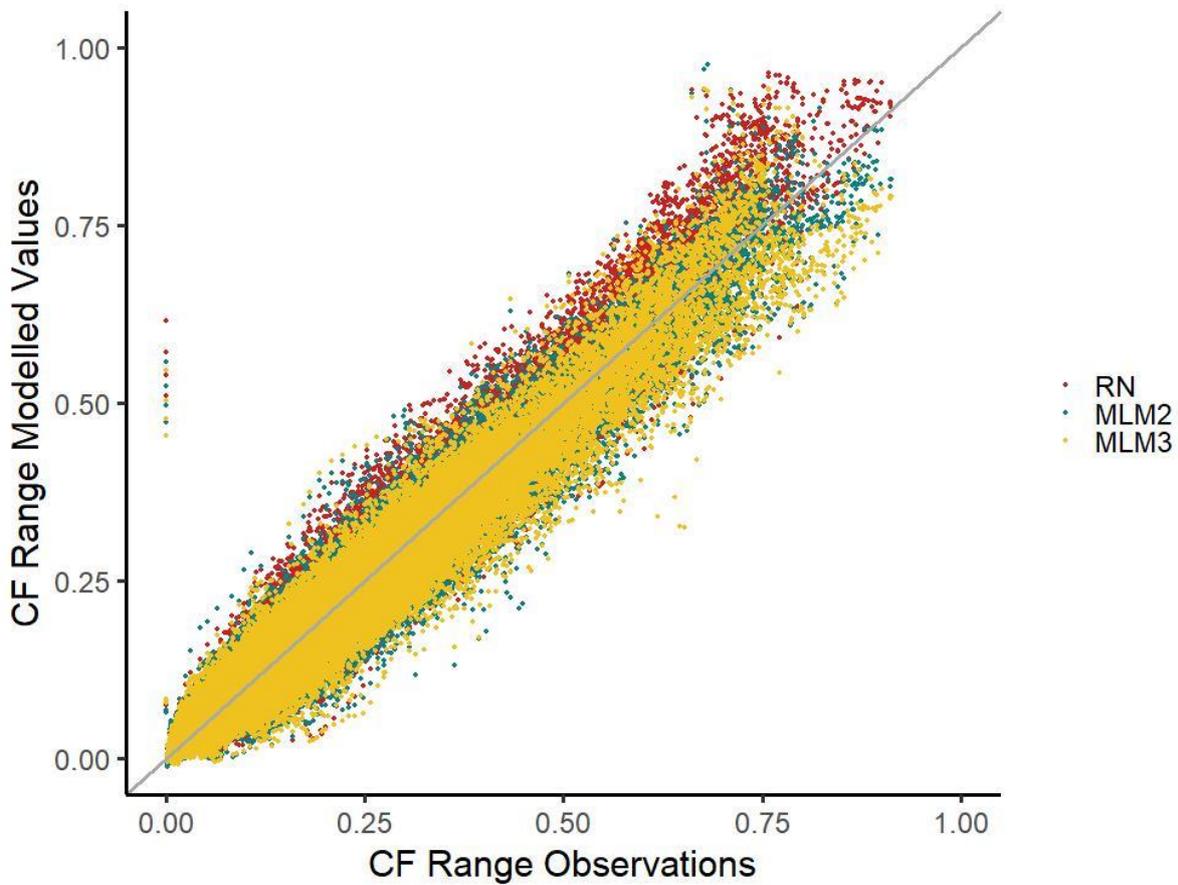

Figure 8: Scatterplot of modelled generation time series compared with observations (Renewables.ninja time series abbreviated as RN and machine learning model time series abbreviated as MLM2 and MLM3 versus observations on the x-axis)

The distributional characteristics of the MLM3 time series are fairly similar to the MLM2 time series. Generation frequencies in the very low CF range (8,254 actual values with a CF < 0.04) are better approximated by the MLM2 time series (7,115 modelled values with a CF < 0.04) than by MLM3 (6,859 modelled values). CF frequencies in the mid-range from around 0.2 to 0.5 (17,772 actual values with a CF between 0.2 and 0.5) are also slightly better approximated by the MLM2 time series (17,397 modelled values between 0.2 and 0.5) than by MLM3 (17,240 modelled values with a CF between 0.2 and 0.5). In the range from 0.1 to 0.2 higher densities can be seen with the MLM3 time series. The same holds true for the range around 0.1 (8,928 actual values between 0.08 and 0.12), where the MLM3 time series (9,300 modelled values) overestimates slightly more than the MLM2 (9,249 modelled values), although not easily visible when comparing densities. In the high CF range (121 observed values > 0.8), the MLM3 (115 modelled values) underestimates and the MLM2 (126 modelled values) overestimates (Figure 9).

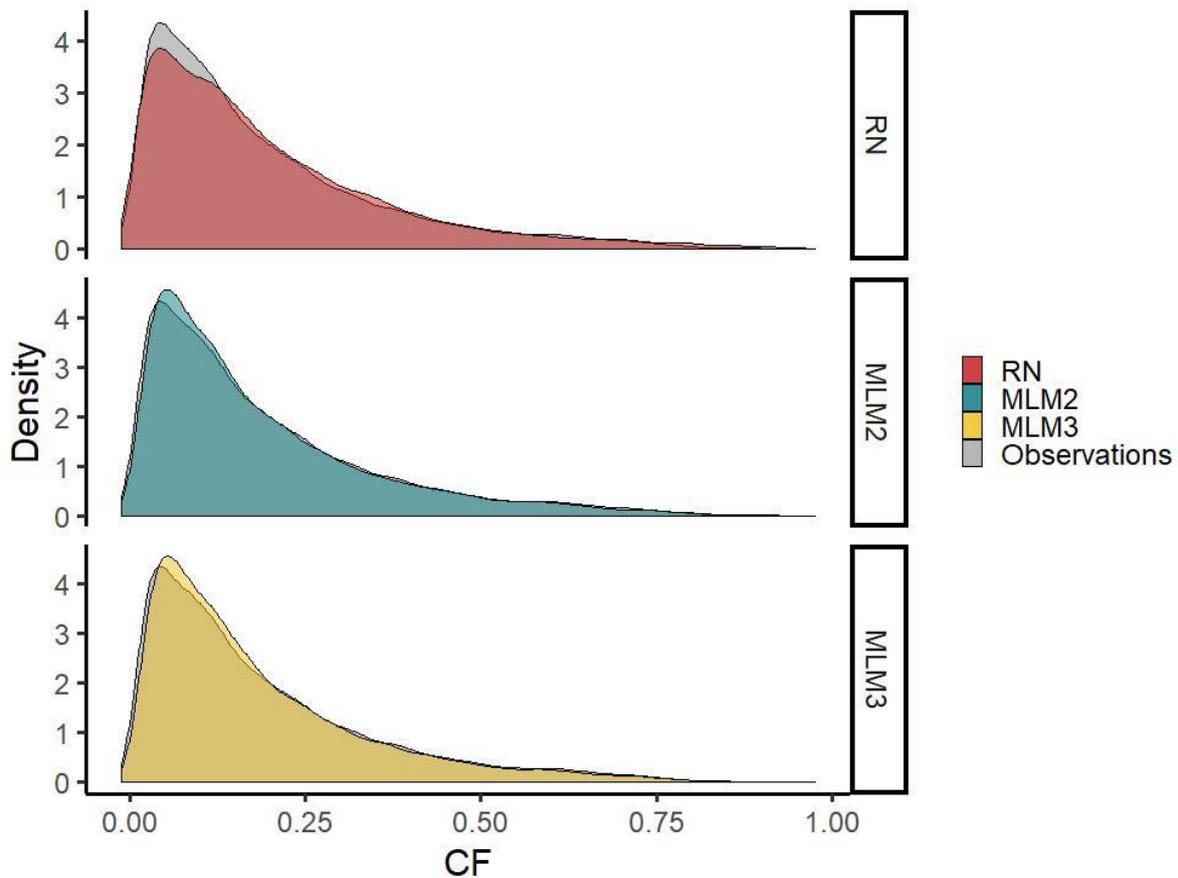

Figure 9: Total distributions of the RN, MLM2 and MLM3 modelled time series compared with observed total distribution (dark coloured areas correspond to an accordance of the modelled and the observed time series; Renewables.ninja time series abbreviated as RN and machine learning model time series abbreviated as MLM2 and MLM3)

### Diurnal Deviations MLM3

The MLM3 median deviations are lower than with the MLM2 in the evening and night (hours 17 - 24) and the morning (hours 1 - 8) hours. Around noon (hours 9 - 14), the MLM2 median deviations are lower.

The differences in the deviation range between the MLM3 and the MLM2 time series do not follow the same patterns as with median deviations, with the MLM3 featuring a narrower deviation range for 8 morning and evening hours (hours 3, 4, 5, 8, 10, 17, 18, 19). The MLM3 also exhibits a deviation range remarkably wider in the early morning hours 2 – 5 than during the remaining hours (Figure 10).

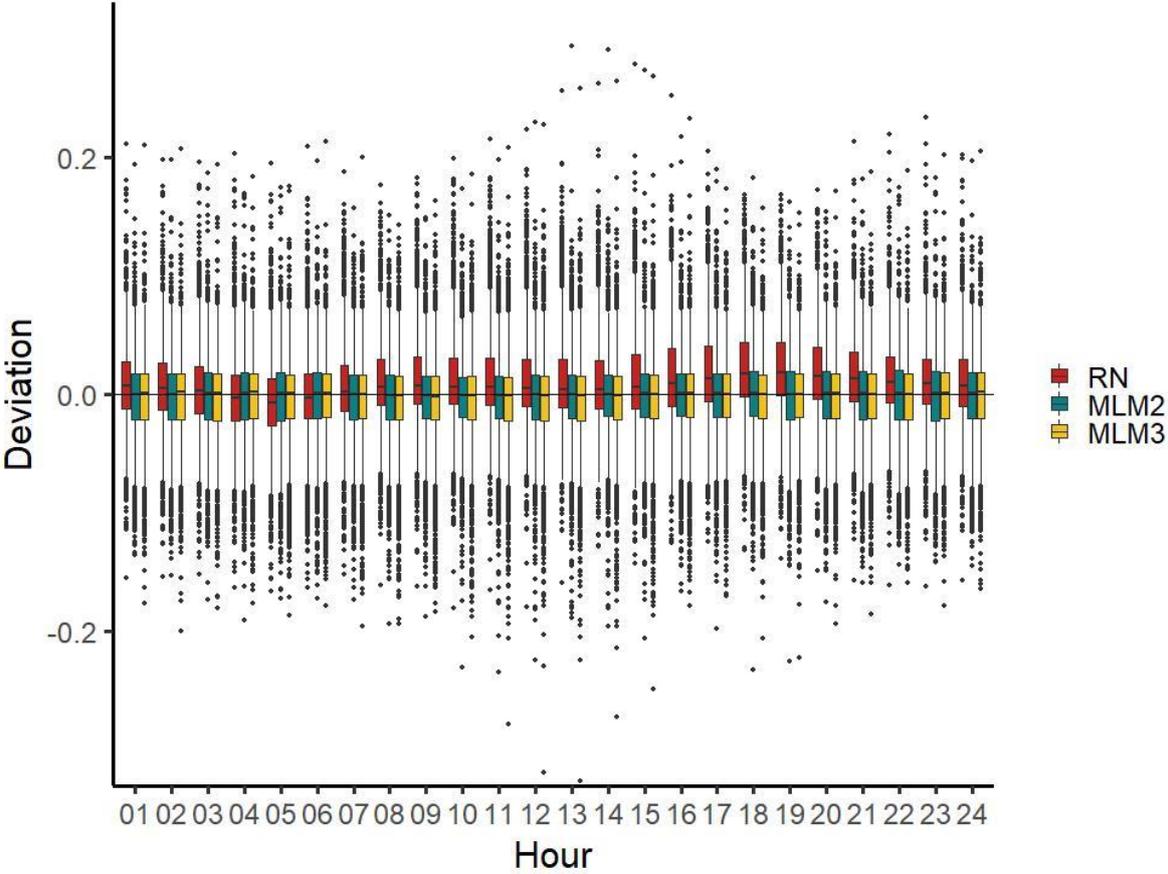

Figure 10: Hourly deviations from observed values for RN, MLM2 and MLM3 (Renewables.ninja time series abbreviated as RN and machine learning model time series abbreviated as MLM2 and MLM3

### Seasonal characteristics MLM3

The median deviation is lower in 8 out of 19 classes and seasons for MLM3 when compared with MLM2, mostly located in the two lowest CF classes (0.0-0.2, 0.2-0.4). The MLM2 time series performs better in 11 out of 19 classes and seasons with CF classes 0-0.2 and 0.6-0.8 being the sweet spot. For both time series a tendency to overestimate events in the lowest CF class can be seen for all seasons. MLM3 works better than MLM2 in 11 out of 19 classes and seasons when comparing deviation ranges mostly within CF classes 0.4-0.6 and 0.6-0.8 (Figure 11).

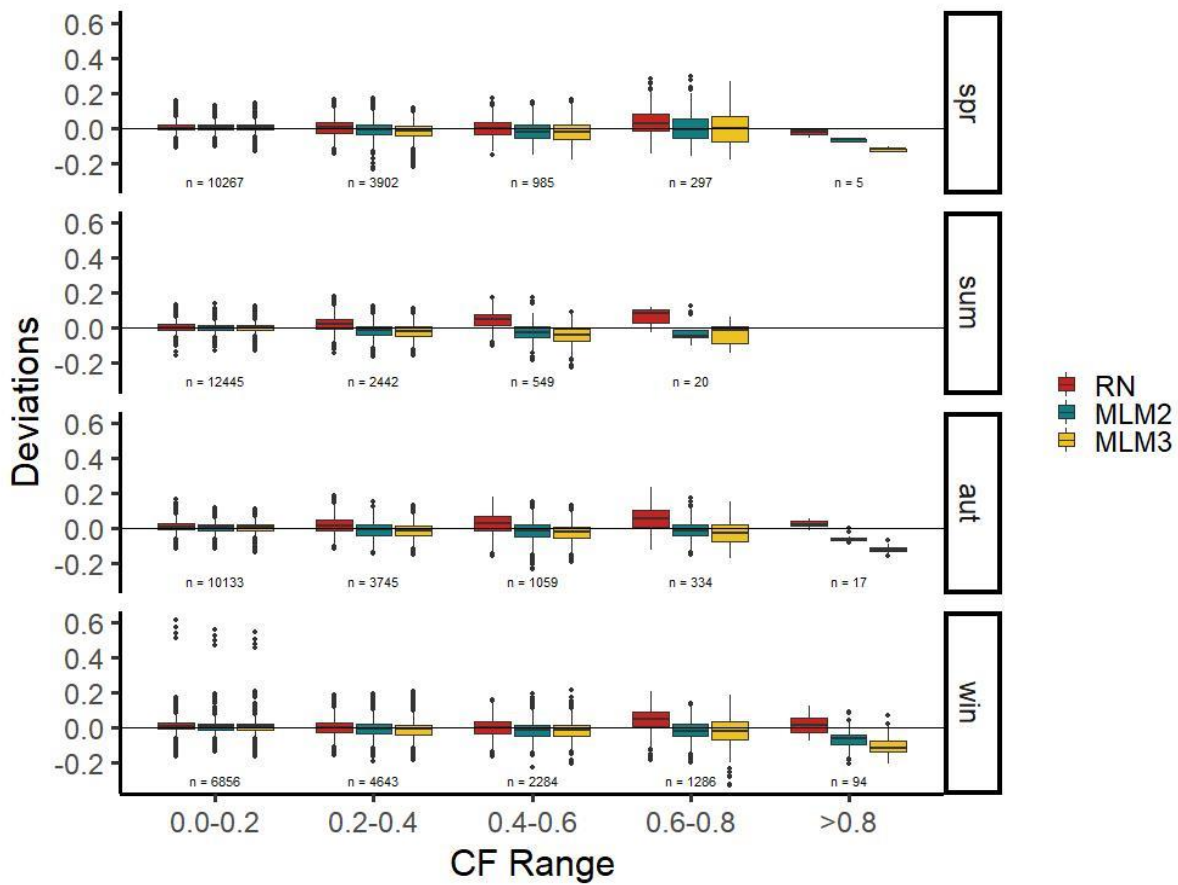

Figure 11: CF deviations of RN, MLM2 and MLM3 modelled time series within different CF bins for separate seasons from 2010 to 2016 (Renewables.ninja time series abbreviated as RN and machine learning model time series abbreviated as MLM2 and MLM3, n denotes the number of observations within capacity factor bins)

### Durations and frequencies of low, high and extreme values MLM3

Frequencies (102 actual events) of low capacity factor extreme values (CF < 0.005) are better approximated by MLM2 (238 modelled events) than by MLM3 (248 modelled events). Mean durations (3.19 consecutive hours of observed generation below 0.005 CF) of low capacity factor extreme values are also better approximated by MLM2 (3.50 consecutive hours of observed generation below 0.005 CF) compared with MLM3 (3.70 consecutive hours of modelled generation below 0.005 CF). The same holds true when low generation extremes maximum durations (10 observed consecutive hours below 0.005 CF) are concerned, where MLM3 time series overestimates maximum duration (15 consecutive hours of modelled generation below 0.005 CF) and MLM2 time series provides an exact match (10 modelled consecutive hours below 0.005 CF). Frequencies (121 actual events) of high capacity factor extreme values (CF > 0.8) are better approximated by MLM2 (125 modelled events) than by MLM3 (115 modelled events). Mean durations of very high generation events (7.56 observed consecutive hours), however, are better approximated by MLM3 (7.19 modelled consecutive hours) than by MLM2 time series (5.21 modelled consecutive hours). Taking the maximum duration into account MLM2 (21 modelled consecutive hours) provides a better estimate MLM3 (17 modelled consecutive hours) (Table 6).

Table 6: Observed and modelled (Renewables.ninja [RN] and machine learning models [MLM1, MLM2]) CF including frequencies, mean and maximum durations of extremely low (<0.005), low (0.01), high (0.75) and extremely high (>0.8) generation events in consecutive hours

| CF Ranges | Observations | RN | MLM2 | MLM3 |
|---|---|---|---|---|
| CF <0.005 | | | | |
| Frequency | 102 | 430 | 238 | 248 |
| Mean Duration | 3.19 | 4.62 | 3.50 | 3.70 |
| Max Duration | 10 | 14 | 10 | 15 |
| CF <0.01 | | | | |
| Frequency | 509 | 748 | 341 | 283 |
| Mean Duration | 2.98 | 2.60 | 1.91 | 1.78 |
| Max Duration | 25 | 12 | 5 | 7 |
| CF >0.75 | | | | |
| Frequency | 204 | 356 | 225 | 187 |
| Mean Duration | 3.46 | 3.10 | 3.41 | 3.46 |
| Max Duration | 13 | 18 | 15 | 21 |
| CF >0.8 | | | | |
| Frequency | 121 | 471 | 125 | 115 |
| Mean Duration | 7.56 | 9.24 | 5.21 | 7.19 |
| Max Duration | 35 | 33 | 21 | 17 |

## Frequencies and ranges of power ramps MLM3

For the shorter time frames (1h, 3h) MLM3 performs better than MLM2 with CF change means and frequencies being approximated better or equal to the MLM2 time series, however MLM2 performs better the longer the time frames (Table 7).

Table 7: Minimum, Maximum, Mean of negative and positive CF ramps within different time frames and frequency of highly positive and negative CF ramps

| Time frame | Observations | RN | MLM2 | MLM3 |
|---|---|---|---|---|
| **1h** | | | | |
| Min | -0.66 | -0.11 | -0.13 | -0.11 |
| Neg. Mean | -0.012 | -0.013 | -0.012 | -0.012 |
| Neg. Freq | 31313 | 32243 | 31132 | 31308 |
| Pos. Mean | 0.013 | 0.014 | 0.013 | 0.012 |
| Pos. Freq | 30054 | 29124 | 30235 | 30059 |
| Max | 0.50 | 0.16 | 0.14 | 0.15 |
| Frequency<-0.2 | 1 | 0 | 0 | 0 |
| Frequency>0.2 | 1 | 0 | 0 | 0 |
| **3h** | | | | |
| Min | -0.67 | -0.28 | -0.33 | -0.30 |
| Neg. Mean | -0.022 | -0.024 | -0.022 | -0.022 |
| Neg. Freq | 94293 | 96677 | 94091 | 94389 |
| Pos. Mean | 0.023 | 0.027 | 0.023 | 0.023 |
| Pos. Freq | 89805 | 87421 | 90007 | 89709 |
| Max | 0.50 | 0.30 | 0.31 | 0.32 |
| Frequency<-0.2 | 75 | 71 | 61 | 69 |
| Frequency>0.2 | 92 | 181 | 93 | 79 |
| **6h** | | | | |
| Min | -0.67 | -0.46 | -0.50 | -0.47 |
| Neg. Mean | -0.035 | -0.038 | -0.034 | -0.034 |
| Neg. Freq | 188949 | 192660 | 188749 | 189180 |
| Pos. Mean | 0.037 | 0.042 | 0.036 | 0.036 |
| Pos. Freq | 179238 | 175527 | 179438 | 179007 |
| Max | 0.51 | 0.50 | 0.47 | 0.47 |
| Frequency<-0.2 | 1,616 | 2,140 | 1,736 | 1682 |
| Frequency>0.2 | 1,846 | 2,921 | 1,860 | 1687 |
| **12h** | | | | |
| Min | -0.68 | -0.60 | -0.64 | -0.62 |
| Neg. Mean | -0.054 | -0.059 | -0.053 | -0.052 |
| Neg. Freq | 376200 | 380753 | 376127 | 377137 |
| Pos. Mean | 0.056 | 0.064 | 0.055 | 0.055 |
| Pos. Freq | 360138 | 355585 | 360211 | 359201 |
| Max | 0.62 | 0.67 | 0.68 | 0.63 |
| Frequency<-0.2 | 14,185 | 18,009 | 14,245 | 13396 |
| Frequency>0.2 | 14,680 | 19,194 | 14,237 | 13580 |